\documentclass[prd,aps,superscriptaddress,floatfix,nofootinbib,eqsecnum]{revtex4-2}

\pdfoutput=1

\usepackage{amsfonts}
\usepackage{amsmath}
\usepackage{amssymb}
\usepackage{bm}
\usepackage{dcolumn}
\usepackage{graphicx}
\usepackage[latin1]{inputenc}
\usepackage{latexsym}
\usepackage{rotating}
\usepackage{hyperref}
\usepackage{subfigure}
\usepackage{color}
\usepackage{changes}
\usepackage{verbatim}
\usepackage{comment}
\usepackage{empheq}
\usepackage{csquotes}
\usepackage{physics}
\usepackage{float}
\usepackage{amsmath,latexsym}
\usepackage{booktabs}  

\begin{document}

\title{Quantum Gravity Corrections to Hawking Radiation via GUP}
\author{Gaurav Bhandari}\email{bhandarigaurav1408@gmail.com}\affiliation{Department of Physics, Lovely Professional University, Phagwara, Punjab, 144411, India}

\author{S. D. Pathak}\email{shankar.23439@lpu.co.in}\affiliation{Department of Physics, Lovely Professional University, Phagwara, Punjab, 144411, India}

\author{Manabendra Sharma}\email{sharma.man@mahidol.ac.th}\affiliation{Centre for Theoretical Physics and Natural Philosophy, Nakhonsawan Studiorum for Advanced Studies, Mahidol University, Nakhonsawan, 60130, Thailand}

\author{Maxim Yu Khlopov}\email{khlopov@apc.in2p3.fr}\affiliation{Virtual Institute of Astroparticle Physics, 75018 Paris, France\\
Institute of Physics, Southern Federal University, Rostov on Don 344090, Russia\\ National Research Nuclear University MEPHI, 115409 Moscow, Russia.}

\begin{abstract}

In this paper we explore the effects of a  Generalized Uncertainty Principle (GUP) on Schwarzschild black hole. In particular, we incorporate the effects of GUP  into the   Parikh-Wilczek tunneling process for Hawking radiation. To this effect, we observe that  results obtained due to GUP correction resemble that of the Reissner-Nordstr\"{om} black hole, showing similarities to the nature of an electric charge. We also find that, within this framework, the emission is not purely thermal, thus addressing the information loss problem through the correlation function.
\end{abstract}

\maketitle



\section{Introduction} \label{Introduction}
The discovery of General Relativity (GR), proposed by Albert Einstein in 1915, provided a mathematical framework to construct many cosmological models as well to explain astrophysical phenomena. 
The profound  success of GR can be seen from prediction like perihelion precession of mercury\cite{rana1}, deflection of light when passing through massive bodies\cite{ge1}, and gravitational redshift of light\cite{woj1}. In addition to this one of the most intriguing predictions of GR is the mysterious supermassive astronomical objects having a very large gravitational attraction that even light cannot escape are Black Holes\cite{haw1,o1,ku1}.

A vast number of studies have been done on black hole formation, properties, radiation, black hole shadows, and so on exist in the literature \cite{cite1,cite2,cite3,cite4,cite5}. General Relativity (GR) provides a framework for understanding these properties of black holes. Black holes may help in understanding the beginning and creation of the universe \cite{sid2,cite6,cite7}.

One of the properties of black holes is that they radiate. A black hole can be thought of as a black body emitting thermal radiation with a temperature $T_H$, known as the Hawking temperature. This Hawking temperature is inversely proportional to the mass $M$ of the black hole. Therefore, due to conservation laws, as the black hole radiates, its mass decreases. This process continues until its mass approaches zero, resulting in a divergence in temperature. It behaves as an ordinary quantum system from an outside observer's view. However, Hawking objected to this idea through the "Hawking information paradox" in 1975 \cite{cite8}, arguing that this radiation causes the loss of information within the black hole, increasing the von Neumann entropy of the universe during black hole formation and evaporation processes.

There are several derivations of the Hawking temperature, but none of them provides a direct explanation for the source of radiation. The theoretical prediction of Hawking radiation stems from quantum effects near the event horizon. One of the most intuitive explanations involves particle-antiparticle creation near the event horizon of the black hole. In this scenario, one particle falls into the black hole, and the other escapes, leading to radiation from the emitted particle. The particle that falls into the black hole carries negative energy, reducing the black hole's mass, while the escaping particle contributes to the observed radiation \cite{cite9}. This mechanism, although intuitive, raises several questions regarding the exact nature and conservation laws involved in the process \cite{cite10,cite11,c12,c13}. 

In 2000, Parikh and Wilczek first integrated the outcomes of quantum mechanics into the framework of General Relativity \cite{c14}. They derived Hawking radiation using radial null geodesics and calculated tunneling probability. They considered particles in a dynamic geometry to apply conservation laws. By considering conservation laws, the horizon shrinks with the radiating particles, leading to non-thermal corrections to the black hole's thermal spectrum. This proposal also manages to address the information loss problem to some extent.

Recent research has proposed solutions to issues related to information loss and non-thermal emission within the framework of quantum gravity \cite{c15,c16,c17,c18,s2}. Quantum gravity theories are particularly crucial near the horizon and singularity of black holes, where phenomena occur at extremely high energies and small length scales. Among the many quantum gravity theories, one prominent approach is the Generalized Uncertainty Principle (GUP) \cite{so1,so2,so3}, which reformulates the well-known Heisenberg Uncertainty Principle (HUP) by incorporating linear and quadratic momentum terms suggested by other quantum gravity theories like Loop Quantum Gravity (LQG) \cite{rovelli1998strings,sharma2019background,garay1995quantum,as1,ha1,sh1},  String Theory (ST) \cite{veneziano1986stringy,witten1996reflections,scardigli1999generalized,gross1988string,amati1989can,yoneya1989interpretation,s3} and Doubly Special Relativity (DSR) \cite{double1, gh1}, thereby implying a minimum observable length \cite{m1,si1,sc1,kempf1994quantum,kempf1997quantum,tawfik2015review,lake2021generalised,PhysRevD.85.104029,ADLER_1999}.

With the success of GUP approaches in explaining many cosmological phenomena and models \cite{L1,L2,L3,L4,L5,L6,L7,L8,L9,bh1,bh2}, this paper considers extending the GUP-correction to the Schwarzschild black hole to explore the role of GUP in determining the Hawking temperature via the tunneling process as suggested by Parikh-Wilczek. In Sec.\ref{1}, we introduce the GUP-corrected Schwarzschild line element and express it in Painlev\'{e}-Gullstrand coordinates to achieve stationarity. In Sec.\ref{2}, we derive the expression for tunneling probability and the Hawking temperature. Sec.\ref{3} concludes the results obtained by considering the GUP approach and compares them with the ordinary case. Finally, we investigate the correlation function between particles emitted from the black hole to address the problem of information loss. 

\section{GUP-corrected Schwarzschild black hole}\label{1}

The geeneral relativity (GR) predicts a singularity in spacetime at the center of a black hole. It is believed that quantum gravitational effect, which can not be neglected at high curvature region, could regulate the occurrence of singularity.  Various approaches have been proposed to address this issue using quantum gravity theories \cite{mod1,mod2,mod3}. In this context, we will use a bottom up approach by invoking a Generalized Uncertainty Principle (GUP)  which involves modifying the Heisenberg Uncertainty Principle (HUP) to obtain a minimal length, as suggested by many candidates of quantum gravity theories. There are many forms of GUP, but the most familiar  one is the following given in \cite{ali8}:
\begin{equation}
\Delta x \Delta p \geq \frac{\hbar}{2}\left(1+\alpha L_P^2\frac{\Delta p^2}{\hbar^2}\right),
\end{equation}
where $\alpha$ is the dimensionless GUP parameter of order unity and $L_p$ is the Planck length. This is straight forward to see that we recover the well-known  HUP when $\alpha \to 0$. This GUP have a minimum bound in the uncertainty in position of 
\begin{equation*}
(\Delta x)_{min}=\sqrt{\alpha}L_P ,
\end{equation*}
for simplicity, we choose value of $\hbar=L_P=1$ from here.

For the classical Schwarzschild black hole line element is given as:
\begin{equation}
ds^2= -f(r)dt^2 + \frac{dr^2}{f(r)}+r^2 d\Omega^2,
\end{equation}
where $f(r)$ is given as $\left(1-\frac{2GM}{r}\right)$.
 With this it can be shown that The GUP corrected Schwarzschild line element takes the form \cite{ho1}:
\begin{equation}\label{gup}
ds^2= -g(r)dt^2 + \frac{dr^2}{g(r)}+r^2 d\Omega^2,
\end{equation}
where 
\begin{equation}\label{event}
\begin{split}
g(r) &= \left(1- \frac{2GM}{r}\right)\left(1+ \frac{\alpha}{4r^2}+\frac{\alpha^2}{8r^4}\right)
\\
& \approx 1-\frac{2M}{r}
+\frac{\alpha}{4r^2} 
 .\end{split}
\end{equation}
One can easily observe that it leads to two event horizons associated with the metric coefficient in Eq.(\ref{gup}), one is the inner horizon and other is outer horizon:
\begin{align}
r_{inner} &= \frac{1}{2}[2M - \sqrt{4M - \alpha}] , \label{eq:inner} \\
r_{outer} &= \frac{1}{2}[2M + \sqrt{4M - \alpha}] , \label{eq:outer}
\end{align}
Thus, the GUP-corrected Schwarzschild metric is given by:
\begin{equation}
ds^2= -\left(1-\frac{2M}{r}+\frac{\alpha}{4r^2}\right)dt_s^2 + \left(1-\frac{2M}{r}+\frac{\alpha}{4r^2}\right)^{-1}dr^2+r^2 d\Omega^2 ,
\end{equation}
 where $t_s$ is the Schwarzschild time. The above line element resembles that of Reissner-Nordstr\"{om} electrically charged black hole.

In 2000, the explanation of Hawking radiation through the tunneling effect was first proposed by Parikh and Wilczek, where the horizon is treated within a dynamical geometry in their paper \cite{c14}. Since tunneling occurs across the horizon, it is necessary to choose the stationary and non-singular coordinates at the horizon.This is achieved by introducing new time coordinates as:
 \begin{equation}
    t= t_s + \int\left[\frac{\sqrt{1-g_{00}}}{g_{00}}\right]dr,
 \end{equation}
 where $t_s$ is the Schwarchild time. With this choice, the new Schwarzschild line element in Painlev\'{e}-Gullstrand coordinates reads
  \begin{equation}\label{pgtrans}
    ds^2=-\left(1-\frac{2M}{r}+\frac{\alpha}{4r^2}\right)dt^2 + 2\left(\sqrt{\frac{2M}{r}-\frac{\alpha}{4r^2}}\right)dtdr+dr^2+r^2 d\Omega^2 .
  \end{equation}
\section{Tunneling process for massless particles}\label{2}
 From the Eq.(\ref{pgtrans}) one can deduce the equation of radial null geodesics as
 \begin{equation}\label{null}
\dot{r} = \pm1-\sqrt{\frac{2M}{r}-\frac{\alpha}{4r^2}},    
 \end{equation}
 where the plus and minus signs correspond to outgoing (ingoing) geodesics as t increases towards the future.
 We consider that just inside the outer event horizon, pair creation takes place, creating virtual particle-antiparticle pairs. The absorption and emission of these virtual particles fluctuate the hole's mass through energy conservation law. During the tunneling process, emission of virtual particle with energy $\omega$ reduces the mass $M$ to $M-\omega$ through conservation law. These fluctuations will also affect the space-time geometry just outside the horizon making the geometry dynamical.

Consequently, Eq.(\ref{pgtrans}) and Eq.(\ref{null}) get modified by considering self-gravitating effects as
\begin{equation}
ds^2=-\left[1-\frac{2(M-\omega)}{r}+\frac{\alpha}{4r^2}\right]dt^2 + 2\left[\sqrt{\frac{2(M-\omega)}{r}-\frac{\alpha}{4r^2}}\right]dtdr+dr^2+r^2 d\Omega^2,
\end{equation}
 and 
 \begin{equation}
  \dot{r} = \pm1-\sqrt{\frac{2(M-\omega)}{r}-\frac{\alpha}{4r^2}}.   
 \end{equation}
 We consider that the virtual particle is at $r_{in}$ initially, which is $\zeta$ (very small quantity with the dimension of length) smaller than the outer horizon radius crosses somewhere at $r_{out}$, which is $\zeta$ greater than the reduced outer horizon.
\begin{align}
r_{in} &= \frac{1}{2}\left[2M+\sqrt{4M^2-\alpha}\right] -\zeta , \\
r_{in} &= \frac{1}{2}\left[2(M-\omega)+\sqrt{4(M-\omega)^2-\alpha}\right] +\zeta .
\end{align}

Considering the outgoing wave traced back toward the outer horizon, its wavelength, as measured by the stationary observer, is extremely blue-shifted. Consequently, the radial wavenumber approaches infinity, making the point particle description valid or justifying the WKB (Wentzel-Kramers-Brillouin) approximation. 

One can use the semiclassical procedure to find the transmission coefficient of tunneling particle as an exponential function of imaginary part of action as
 \begin{equation}\label{proba}
\Gamma \approx e^{-2 Im S}.
 \end{equation}

 The imaginary part of action for positive energy particle crosses the horizon $r_{in}$ to $r_{out}$ is expressed as
 \begin{equation}
    \text{Im} S = \text{Im} \int_{r_{in}}^{r_{out}} p_rdr ,
 \end{equation}
 where $p_r$ is the canonical momentum of a virtual particle with Hamiltonian as $H=M-\omega$. And, we can write it as
 \begin{equation}
    \text{Im} S = \int_{0}^{\omega}\int_{r_{in}}^{r_{out}} \frac{dr(-d\omega)}{\dot{r}} .
 \end{equation}
 This integral is solved by deforming the contour using \cite{mi1,mi2},
 \begin{equation}
    \text{Im} S = \int_{0}^{\omega} \frac{2\pi\{2(M-\omega)+\sqrt{4(M-\omega)^2-\alpha}\}}{4(M-\omega)\{2(M-\omega)+\sqrt{4(M-\omega)^2-\alpha}\}-2\alpha},
 \end{equation}
finally, the approximate integral is expressed as
\begin{equation}\label{Im}
    \text{Im} S \simeq 2 \pi [2\omega\left(M-\frac{\omega}{2}\right)+\frac{M}{2}\sqrt{4M^2-\alpha}-\frac{M-\omega}{2}\sqrt{4(M-\omega)^2-\alpha}].
\end{equation}
A similar result can be derived for the ingoing radial motion of a virtual particle with negative energy created outside the black hole and tunneling into it. Now, substituting Eq.(\ref{Im}) on Eq.(\ref{proba}) gives
\begin{equation}\label{final}
\Gamma \approx \exp\left\{-4 \pi \left[2\omega\left(M-\frac{\omega}{2}\right)+\frac{M}{2}\sqrt{4M^2-\alpha}-\frac{M-\omega}{2}\sqrt{4(M-\omega)^2-\alpha}\right]\right\} = e^{+\Delta S_{BH}} ,
\end{equation}
where $\Delta S_{BH}$ is the change in Bekenstein-Hawking entropy.
On comparing the power of exponential term in Eq.(\ref{final}) to the thermal emission $e^{-\beta \omega}$, $T_{H}$ at Hawking temperature  only upto the first order in $\omega$ gives
\begin{equation}\label{haw}
    T_H= \frac{\sqrt{4M^2-\alpha}}{2\pi[2M+\sqrt{4M^2-\alpha}]^2}.
\end{equation}
The Hawking temprature we obtained in Eq.(\ref{haw}) resemblesthe Hawking temperature obtained in  the Reissner-Nordstr\"{om} black hole as 
\begin{equation}\label{eq}
   \alpha = Q^2
\end{equation}
The quantum effects on a Schwarzschild black hole behave similarly to those on an electrically charged black hole. This analogy leads us to conclude that electric charge behaves similarly to quantum effects. Such studies have already been conducted in a non-commutative background \cite{no1}. This implies that the existence of electric charge in spacetime leads to quantum fluctuations in the background.
As the GUP parameter is related to the electric charge in Eq.(\ref{eq}), we have a lower bound on choosing this GUP parameter as Q is quantized as $\alpha=Q^2= 2.56 \cross 10^{-38}$  indicating a situation akin to minimal length, near the Planck length.

\begin{figure}[H]
    \centering
    \includegraphics[width=0.6\textwidth, height=0.3\textheight]{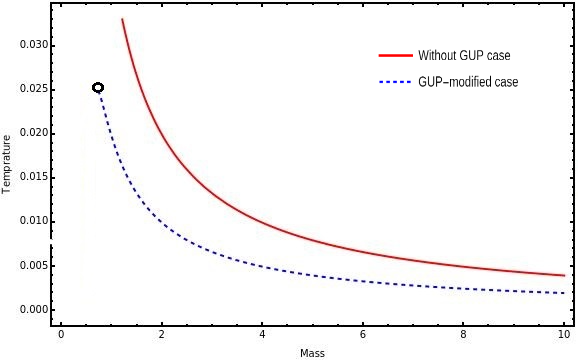}
    \caption{Graph between temperature and mass of black hole where red line depicts the ordinary case obtained by Parik-Wilczek and the blue dash line represents the GUP-deformed case where $\alpha=1.5$. }
    \label{fi1}
\end{figure}
From Fig.(\ref{fi1}) we depict that for the ordinary case of black hole evaporation, the process terminates with the black hole reaching an infinitely high temperature, ultimately leading to the complete disappearance of the black hole without leaving any remnant behind. For the GUP-deformed scenario, black hole evaporation continues until the black hole reaches a finite mass, leaving behind a remnant with a radius $r=a$. In this scenario, $a$ is the minimal length parameter, the smallest GUP-deformed non-singular Schwarzschild black hole radius. Consequently, this GUP deformation stops the further evaporation of black holes.
\section{Conclusions}\label{3}
We have derived the Hawking temperature using a Parik-Wilczek tunneling mechanism within a more viable framework that incorporates quantum effects through one of the quantum gravitational approaches, known as GUP. First, we chose the GUP-deformed Schwarzschild line element and then transformed this line element by introducing a new time coordinate to render the spacetime stationary and non-singular. We found that after incorporating quantum gravitational effects via GUP, the Hawking temperature is modified from the original obtained by Parikh-Wilczek. We observe from Eq.(\ref{haw}) the modified Hawking temperature contains new term $\alpha$ represents quantum gravitational parameter.

We obtained the following conclusion based on Fig.(\ref{fi1}) and Eq.(\ref{haw}):

In the GUP-deformed scenerio, Schwarschild black hole does not evaporate completely but leaves a stable remnant with finite mass and with maximum temperature which is opposite to the without GUP case where the black hole evaporates completely. The presence of a minimal length scale introduced by GUP leads to the lower bond on the black hole mass, preventing it to became zero.
In the ordinary case, the temperature of a black hole diverges to infinity as it completely evaporates. However, with the Generalized Uncertainty Principle (GUP), the temperature reaches a maximum value, leaving behind a finite remnant.

 We depict from the modified Hawking temperature under GUP predict a non-thermal emission spectrum due to quantum corrections. The finite temperature during the final stages of black hole evaporation suggests that the emitted radiation contains information, potentially resolving the information loss paradox. The GUP corrections leads to address the information loss problem can be shown through the correlation function between the emitted  particles as
 \begin{equation}
    \chi(\omega_1+\omega_2;\omega_1,\omega_2) \equiv \ln[{\Gamma(\omega_1+\omega_2)}]-\ln[\Gamma(\omega_1)\Gamma(\omega_2)], 
 \end{equation}
 where $\omega_1$ and $\omega_2$ are energies of two emitted particles.
\begin{equation}
\begin{aligned}
\chi(\omega_1+\omega_2;\omega_1,\omega_2) = 4\pi\Bigg[&2\omega_1\left(M-\frac{\omega_1}{2}\right) + 2\omega_2\left(M-\frac{\omega_2}{2}\right) - 2(\omega_1+\omega_2)\left(M-\frac{\omega_1+\omega_2}{2}\right) \\
&-\frac{M-\omega_1}{2}\sqrt{4(M-\omega_1)^2-\alpha} - \frac{M-\omega_2}{2}\sqrt{4(M-\omega_2)^2-\alpha} \\
&+ \frac{M-(\omega_1+\omega_2)}{2}\sqrt{4(M-(\omega_1+\omega_2))^2-\alpha} + \frac{M}{2}\sqrt{4M^2-\alpha}\Bigg],
\end{aligned}
\end{equation}
since the value of the correlation function is nonzero therefore the radiation is not thermal emission. The tunneling probability of two particles having energy $\omega_1$ and $\omega_2$ is not the same as the tunneling probability of a particle with total energy $(\omega_1+\omega_2)$. This indicates that there is a correlation between two emitted particles implying that information about the black hole can be encoded in these correlations.


\begin{acknowledgments}
The research by M.K. was carried out in Southern Federal University with financial support from the Ministry of Science and Higher Education of the Russian Federation (State contract GZ0110/23-10-IF).
\end{acknowledgments}



\end{document}